\documentclass[11pt,a4paper]{article}%\documentclass[bibliography=totocnumbered,listof=totocnumbered,openany,12pt]{scrbook}
\usepackage[makeroom]{cancel}
\usepackage{bm}
\usepackage{amsmath}
\usepackage{amsfonts}
\usepackage{amssymb}
\usepackage{graphicx}
\usepackage{authblk}
\usepackage{cite}
\usepackage[left=2.4cm,top=2.4cm,right=2.4cm,bottom=2.4cm,nohead]{geometry}
\DeclareGraphicsExtensions{.png,.jpg,.pdf}
\usepackage{epstopdf}
\usepackage{float}
\usepackage{fouriernc} 
\usepackage{enumitem}
\usepackage[utf8]{inputenc}
\usepackage[english]{babel}
\usepackage{mathtools}
\usepackage{amscd}
\usepackage{mathrsfs}
\usepackage{amsthm}
\usepackage{siunitx}%For SI units
\usepackage{caption}
\usepackage{subcaption}
\setlist{nolistsep,leftmargin=*}

\usepackage{mhchem} 
\usepackage{cleveref} 
\usepackage{xcolor}
\usepackage{soul}
\usepackage{mdframed}
\usepackage{framed}
\usepackage{longtable}
\DeclareMathAlphabet{\mathpzc}{OT1}{pzc}{m}{it}

\title{Analysis of concentration polarization in reverse osmosis and nanofiltration: zero-, one-, and two-dimensional models}

\makeatletter

\renewcommand\AB@authnote[1]{\textsuperscript{\normalfont#1}}

\makeatother

\author[1]{P.M. Biesheuvel,}
\author[2]{S.~Porada,}
\author[3]{B.~Blankert,}
\author[4,5]{I. Ryzhkov,}
\author[6]{M. Elimelech}

\affil[1]{Wetsus, European Centre of Excellence for Sustainable Water Technology,  %Leeuwarden, 
The~Netherlands.}
\affil[2]{Department of Process Engineering and Technology of Polymer and Carbon Materials, Wroclaw University of Science and Technology, %Wyb. St. Wyspiańskiego 27, 50-370 Wrocław, 
Poland.}
\affil[3]{Water Desalination and Reuse Center, % (WDRC), Biological and Environmental Science and 6 Engineering Division (BESE), 
King Abdullah University of Science and Technology, % (KAUST), 7 Thuwal 23955-6900, 
Saudi Arabia.}
\affil[4]{Institute of Computational Modeling SB RAS, %Akademgorodok 50, 660036 
Krasnoyarsk, Russia.}
\affil[5]{Siberian Federal University, %Svobodny 79, 660041, 
Krasnoyarsk, Russia.}
\affil[6]{Department of Chemical and Environmental Engineering, Yale University, %New Haven CT%06520-8286
USA.}

\date{} 

\newcommand{\s}[1]{\mathrm{_{#1}}}
\newcommand{\kd}{\left\langle \vphantom{k^\text{xy}}\right. \! k^\text{d} \!  \left.\vphantom{k^\text{xy}}\right\rangle}
\begin{document}

\maketitle

\begin{abstract}

Reverse osmosis and nanofiltration are membrane-based methods that remove solutes from solvent, for instance they remove salts from water (desalination). In these methods, an applied pressure is the driving force for solvent to pass the membrane, while most of the solutes are blocked. Very important in the theory of mass transport is the concentration polarization layer (CP layer), which develops on the upstream side of the membrane. Because of the CP layer, the solvent flux through the membrane is reduced while leakage of solutes through the membrane increases, and both these effects must be minimized. So it is very important to understand and describe the nature of the CP layer accurately, especially to find a good estimate of the CP layer mass transfer coefficient, $k$. This is also important for the accurate characterization of membranes in a test cell geometry. 

We theoretically analyse the structure of the CP layer using three levels of mathematical models. First, we present a modification of an equation for $k$ by Sherwood \textit{et al}.~(1965) and show that it works very well in a zero dimensional model. Second, we evaluate a one-dimensional model that is more accurate, which can incorporate any equation for the flow of solvent and solutes through the membrane, and which also makes use of the new modified Sherwood equation. Finally, we fully resolve the complete channel in a two-dimensional geometry, to validate the lower-order models and to illustrate the structure of the CP layer. The overall conclusion is that for typical test cell conditions, the modified Sherwood equation can be used to characterize the CP layer, also when solvent flux through the membrane changes between inlet and outlet of the test cell. Furthermore, the one-dimensional model accurately describes solute removal (for instance water desalination) not just in a short test cell but also in a longer module.

\end{abstract} 

\section{Introduction}

In the present work, we extend the theory of concentration polarization in reverse osmosis (RO) and nanofiltration (NF) membrane modules from the stirred cell geometry discussed in ref.~\cite{CP_arxiv_2023} (which we call `part~I'), to the two-dimensional (2D) geometry of a flow channel. We describe here the high pressure feed channel of an RO unit, with a planar membrane lining one side of the channel.\footnote{Some calculations are based on a channel with a membrane lining both sides of the channel.} We focus on the situation that the flow channel is open, i.e., there is no spacer material. Because of an applied pressure, solvent (in many cases water) flows across the membrane to a permeate channel where it is collected as product. Also some of the solutes in the feedwater (such as salt ions) pass the membrane, and their passage must be minimized. Water and salts that do not pass the membrane eventually flow out of the feed channel and are collected as retentate (also called concentrate or brine). The fraction of water that passes the membrane is the water recovery (ratio). 

Because water flows across the membrane, a concentration polarization layer (CP layer) is formed on the feed side of the membrane. Because of this CP layer, the salt concentration there, at the membrane surface, increases by a certain amount, which can be 30\% or 50\%, but also a factor 2 or more is possible. This increase in salt concentration at the channel/membrane interface reduces the flow of water across the membrane because of the increase in osmotic pressure. At the same time this higher concentration leads to more leakage of salts and other solutes across the membrane, and thus the rejection of the membrane decreases~\cite{Dresner_1964, Brian_1965, Oren_2021}. In the calculation that was discussed in part I, this layer is the same all over the membrane surface, and can therefore be described by a solute mass balance that only considers changes in the direction to/away from the membrane. This mass balance includes transport by diffusion and convection and after integration results in an equation for the structure of the CP-layer that is called the exponential law. This exponential law is arrived at both in a standard model with a fixed film layer thickness across which all solutes must flow between bulk and membrane, and in a more advanced model that incorporates solute flow along the membrane as a `source/sink'-term. This is a good approach when the feed side of the test cell is stirred. However, this is generally not the case, and instead in most test cells and commercial modules the structure of the CP layer changes through the cell or module from inlet to outlet, and thus the exponential law is not expected to be valid. For these cell and module geometries where the feed is not stirred, we must find correct models for the CP effect~\cite{Rohlfs_2016}, especially for the coefficient of solute mass transfer in the CP layer, \textit{k}. We need these models because otherwise we cannot accurately predict the performance of a membrane module, and when a cell is used to test a membrane, we might make an error in the characterization of a membrane.

In this work, we evaluate flow in the feed channel of a pressure-driven membrane process for RO or NF, and present mathematical results at several levels of complexity (0D, 1D, and 2D). First, we present an analytical result for \textit{k} which is based on literature~\cite{Sherwood_1965}, but we improve it. We analyse the improved equation and show that for test cell analysis it is a highly accurate result. However, there is a technical complication relating to the choice of how to average over the membrane surface, which depends on the equations for transport of salts and solvent across the membrane. This problem in the averaging procedure disappears %is resolved 
when we use a 1D model that describes flow of salt and solvent through the feed channel (from inlet to outlet along the membrane). In this 1D model, we do not need to average \textit{k}, and thus we can evaluate any type of equation for membrane transport without a change to the model for the CP layer. Finally we present results of a full 2D model for flow of solvent and salt, which can also be combined with any set of equations for membrane transport. This full 2D model describes flow in the feed channel without having to add a separate CP-model.  

The theory is valid for a solution with any number of neutral solutes, as well as for a binary salt solution (i.e., a solution with one cation and one anion). It also applies to mixtures of neutral solutes and a binary salt solution. But when we have three or more ions with different diffusion coefficients or membrane transport relations, we need a more detailed analysis. In part I it was concluded that ion activity effects in solution are small and can be incorporated by changing \textit{k} slightly. Therefore we do not consider these ion activity effects in the present work in the description of diffusion. However, we do include a correction to the osmotic pressure because of ion activities.

\section{The modified Sherwood-Brian-Fisher-Dresner equation for concentration polarization in pressure-driven membrane flow}
\label{section_0D}

For the CP layer that develops in a channel with membranes on the side walls, most theoretical literature is based on solution of the Lévêque-Graetz (LG) problem. The LG problem describes convective flow through a rectangular channel in the axial, i.e., longitudinal, \textit{z}-direction, with `sideways' diffusion of mass or heat towards the walls (where in our case membranes are located), with a fixed concentration (or temperature) on the side walls. This problem neglects convection of solutes towards the side walls (membrane), so in that direction, mass (heat) transfer is by diffusion (conduction) only. The solute molar flux through the membrane is calculated from evaluating $ J\s{s}\left(z\right) = - D \cdot {\partial c }/{ \partial x }$ at the membrane surface (or for heat transfer, $J\s{h}\left(z\right) = - \lambda \cdot {\partial T }/{ \partial x } $), and then a mass transfer coefficient is defined as $k^\text{d} \left( z \right) = J\s{s}\left( z \right) / \left( c_\infty - c\s{m} \right)$, where $c_\infty$ is the concentration outside the CP layer, and $c\s{m}$ the concentration at the membrane, sometimes called interface concentration, $c\s{int}$. Here we add index `d' to the \textit{k}-factor to indicate this is a mass transfer coefficient defined for a purely diffusional process, different from the convection-diffusion problem we discuss further on in this paper. The solute molar flux $J\s{s}$ and transfer coefficient ${k}^\text{d}$ will be \textit{z}-dependent, but $c_\infty$ and $c\s{m}$ are constants in this analysis. The solution of the LG problem is that the local mass transfer coefficient, $k$, is given by~\cite{Zydney_1986,De_1997, Kim_2005,Zydney_1997}
\begin{equation}
k = \alpha \cdot \left( \frac{D^2 \gamma }{ z}\right)^{1/3} 
\label{eq_k_definition}
\end{equation}
%
% above on purpose ^d left out because this formula will also be used for the convecton+diffusion problem
%
where \textit{z} is the coordinate directed through the channel from inlet to outlet, with position $z \! = \! 0$ the entrance of the channel, i.e., the upstream edge of the membrane. In the LG problem, the prefactor in Eq.~\eqref{eq_k_definition} is $\alpha \! = \! 0.538$.\footnote{This number is equal to $\alpha = 1 / \left( 3^{1/3} \cdot \int_0^\infty \exp \left(-\eta^3 /3 \right)  \text{d}\eta \right)$, \cite{Zydney_1986,De_1997, Kim_2005}. It is also the same as the factor $1.119\cdot 9^{-1/3}$ in Eq.~(15) in ref.~\cite{Zydney_1997}.} The diffusion coefficient is \textit{D}, and for a binary salt solution the harmonic mean diffusion coefficient is used for \textit{D}. The shear rate $\gamma$ is the gradient in fluid axial velocity, $v_z$, in \textit{x}-direction, where \textit{x} is a coordinate in the direction to the membrane, evaluated at the membrane surface. For a parabolic profile of the axial velocity, this shear rate at the membrane surface is $\gamma = 6 \, U / H$, where \textit{U} is the crossflow velocity, which is the water flow rate (in m\textsuperscript{3}/s) through the channel, divided by the cross-sectional area (which is channel width \textit{W} times channel height \textit{H}). We can also call it the average water velocity in \textit{z}-direction, $\left\langle v_z \right\rangle$. %The mass transfer coefficient ${k}^\text{d}$ can also be written as $k_\s{CP}$ or $k\s{dbl}$ (with dbl an abbreviation for the diffusion boundary layer). 
The unit of $k$ is length per time, which can be expressed as m/s, but also as L/m\textsuperscript{2}/hr, which is often written as LMH. 

Now, this analysis has hardly any relation to the RO and NF problem where we have convection of solutes to the membrane and an (almost) equal backdiffusion, so the net transport of solutes through the wall is small. Instead, in the LG problem all of the diffusion to the wall results in removal through the wall. Still, the correlation for $k$ provided in Eq.~\eqref{eq_k_definition} is used in calculations of RO systems, but is now implemented in the exponential law that was discussed in Part I, such as for perfect rejection of solutes given by
\begin{equation}
c\s{m}\left(z\right) = c_\infty \left(z\right) \cdot \exp\left({J\left(z\right)} / {k\left(z\right) }  \right)
\label{eq_exponential_law}
\end{equation} 
where $J$ is the flux or velocity of solvent through the membrane and $k$ is the mass transfer coefficient for the combined convection-diffusion problem. However, it is actually not certain at all whether this ad-hoc approach --to equate the \textit{k} required in this problem to the $k^\text{d}$ of the diffusion problem above-- should be valid. 

Instead of the LG problem that assumes a fixed wall concentration, for the description of solute flow in RO the boundary condition at the membrane is a relation between the molar flux of solute, $J\s{s}$, the volumetric flux of solvent, $J$, and membrane concentration, $c\s{m}$, with all these factors \textit{z}-dependent.\footnote{This transmembrane solvent flux or solvent velocity, $J$, or alternatively called water flux, filtrate flux, or permeate flux, can also be written as $J\s{v}$ with `v' for volume, as $J\s{w}$, with `w' for water, or written as $v\s{w}$. It is also written as \textit{V}.} For a fixed flow of solvent through the membrane and 100\% solute rejection, the boundary condition that relates the gradient ${\partial c } / { \partial x }$ at the membrane to $c\s{m}$, is $c\s{m} \cdot J = D \cdot \left. {\partial c } / { \partial x }\right|_\text{m}$ where coordinate \textit{x} points toward the membrane. This problem of flow of solvent and solutes through an infinitely wide rectangular channel with fixed solvent flux through side walls (and 100\% rejection of solutes) was extensively discussed in Sherwood \textit{et al}. (1965)~\cite{Sherwood_1965}. This same material is also discussed at length in Probstein (1989)~\cite{Probstein_1989}. A related paper with variable fluxes \textit{J} and $J\s{s}$ was also published in 1965 by Brian, which we will discuss later~\cite{Brian_1965}. Recently Johnston \textit{et al}.~\cite{Johnston_2023} also made this calculation for a channel with a single membrane, including a solvent flux that depended on local osmotic pressure (results in their Fig.~7). 

\begin{figure}
\centering
\includegraphics[width=0.5\textwidth]{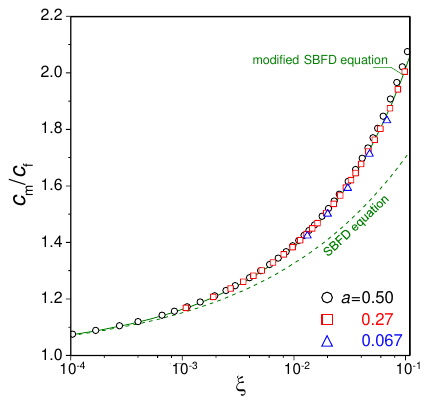}
\vspace{-5pt}
\caption{Membrane concentration, $c\s{m}$, as function of dimensionless position in membrane module, $\xi$, for three values of $a=D/Jh$, where $h$ is the channel half height. Symbols are results of 2D numerical calculations, the dashed line is the SBFD equation~\cite{Sherwood_1965}, and the solid line is the modified SBFD equation, based on Eqs.~\eqref{eq_k_definition} and~\eqref{eq_exponential_law} with $\alpha=0.662$.}
\label{fig_cmcf}
\end{figure}

The key result of Sherwood \textit{et al}. that we analyse, is an expression for $\Gamma$, which is defined as $\Gamma = c\s{m}/c\s{m.c.} \! - \! 1$, where $c\s{m.c.}$ is the `mixed cup' concentration, which is a $z$-position-dependent concentration calculated from dividing the total molar flow of solutes in \textit{z}-direction by the volume flow rate in that direction, given by
\begin{equation}
c\s{m.c.}\left(z\right) = \frac{\int_0^H v_z\left(x,z\right)c\left(x,z\right) \text{d} x}{\int_0^H v_z\left(x,z\right) \text{d} x}
\end{equation}
which in case the concentration is assumed to be invariant across the channel height, i.e., in \textit{x}-direction, as we will assume in the 1D model explained further on, simplifies to $c\s{m.c.} = c$. Results are presented as function of $\xi$, which is a dimensionless \textit{z}-coordinate, defined as $\xi = J^3 z / \gamma D^2$. For $\xi \! < \!  0.02$, Sherwood \textit{et al}. derive the result $\Gamma = \beta \,\xi^{1/3} $, where $\beta \! = \! 1.536$. For these conditions $c\s{m.c.} \! \sim \! c\s{f}$, and thus this equation can be translated to $c\s{m}/c\s{f}=1+\beta \,\xi^{1/3}$. We call this the Sherwood-Brian-Fisher-Dresner (SBFD) equation, but it is not much in agreement with the full 2D numerical results that we discuss in section~\ref{section_2D} (that agree with numerical results of Sherwood \textit{et al}.), see dashed line in Fig.~\ref{fig_cmcf}. Instead, a very accurate function which we propose without a theoretical derivation is a modified SBFD equation, $c\s{m}/c\s{f} = \exp\left( \beta \xi^{1/3} \right)$, see the solid line in Fig.~\ref{fig_cmcf}. We use Eq.~\eqref{eq_exponential_law} to convert the new expression for $c\s{m}/c\s{f}$ to $k$, and then we arrive exactly at Eq.~\eqref{eq_k_definition}, with $\alpha = 1 / \beta$. For a very close fit to full 2D numerical results, see Fig.~\ref{fig_cmcf}, we reduce $\beta$ slightly, to $\beta \! = \! 1.51$ and then we have $\alpha \! = \! 0.662$. Thus, in Fig.~\ref{fig_cmcf} these two equations are compared with full 2D numerical calculations where membrane concentration $c\s{m}$ (divided by feed concentration $c\s{f}$) is plotted against dimensionless position, $\xi$, for three values of a dimensionless parameter \textit{a} which is $a=D/Jh$ where $h$ is the channel half height.\footnote{These specific values of \textit{a} are chosen because for these values Sherwood \textit{et al}.~\cite{Sherwood_1965} present calculation results.}\textsuperscript{,}\footnote{Interestingly, in this 2D calculation, the problem is fully defined when $\xi$ is known as well as $a$ (when we have a fixed permeate flux \textit{J} and 100\% solute rejection).} 

Thus, the modified SBFD equation is a new expression for \textit{k} that has the same dependence on $\gamma$, \textit{D}, and $z$ as the classical LG-problem, but only the prefactor is different. It is quite remarkable that also in the modified SBFD equation \textit{k} is not a function of the permeate flux, \textit{J}. This is also very important because it suggest that in the 1D model that we will discuss in the next section, we can use the modified SBFD equation for the local (i.e., \textit{z}-dependent) \textit{k} even when solvent flux changes with \textit{z} and the exact theory no longer applies. 

But first we discuss the result of averaging \textit{k} over the surface of the membrane, i.e., averaging from $z \! = \! 0$ to $z \! = \! L$. In the LG-problem, because of the constant membrane concentration (and we also assume constant bulk concentration), an average solute (heat) flux can be derived by integrating \textit{k} from $z\!=\!0$ to $z\!=\!L$, and then one arrives at
\begin{equation}
\kd = L^{-1} \int_0^L k \text{d}z = \left\langle \alpha \right\rangle \left(\frac{D^2 \gamma}{L}   \right)^{1/3}
\label{eq__avg_LG}
\end{equation}
where $\left\langle \alpha \right\rangle = \tfrac{3}{2} \, \alpha = 0.807$~\cite{Zydney_1986,Kim_2005}.\footnote{In literature one often finds a factor 1.85 in relations involving the Sh, Re and Sc numbers as well as the hydraulic diameter. We go from 0.807 to 1.85 by multiplying with $\sqrt[3]{12}$.} However, for the RO problem of combined diffusion and convection to the membrane, this averaging has no meaning. Instead, the averaging that we need depends on how solvent and solutes flow across the membrane, i.e., what equations describe membrane transport. When solute flux is linearly related to $c\s{m}$, a membrane concentration that is averaged over the membrane surface (i.e., over the length of the module), $\left\langle c\s{m} \right\rangle$, is a useful property, given by $\left\langle c\s{m} \right\rangle = L^{-1} \int_0^L c\s{m} \text{d}z$. This is the case because this average membrane concentration will then also be used in the averaged membrane transport equations. The same is the case for solvent flux that via osmotic pressure is also (approximately) linearly dependent on concentration $c\s{m}$. However, when solute transport across the membrane depends on $c\s{m}$ in a different way, for instance dependent primarily on the square of membrane concentration, then another average is more relevant to evaluate, in this example that would be ${\left\langle c_\text{m}^2 \right\rangle}$. 

In the most simple situation the solute flux through the membrane is proportional to $c\s{m}$, and then it is useful to calculate an average $c\s{m}$ according to
\begin{equation}
\frac{\left\langle c\s{m} \right\rangle}{c\s{f}} = \frac{1}{L} \int_0^L \frac{c\s{m}\left(z\right)}{c\s{f}} \text{d}z = \frac{1}{X} \int_0^{X}  \exp\left\{ \alpha^{-1} \xi^{1/3} \right\} \text{d} \xi = {3 } \cdot \left(\frac{1}{p}- \frac{2}{p^2}+ \frac{2}{p^3} \right) \cdot \exp \left( p \right)-\frac{6 }{p^3}
\label{eq_cm_average}
\end{equation}
where we implemented Eqs.~\eqref{eq_k_definition} and~\eqref{eq_exponential_law}, and we introduce the shorthand notation $p=\sqrt[3]{X} \, / \, \alpha$. The dimensionless coordinate is $\xi = z / z\s{ref}$, with a reference length $z\s{ref} = D^2 \gamma / J^3$, and dimensionless module length $X = L / z\s{ref}$. Based on a series expansion, Eq.~\eqref{eq_cm_average} can be approximated as
\begin{equation}
\frac{\left\langle c\s{m} \right\rangle}{c\s{f}} \sim \exp\left(\tfrac{3}{4} \, p \right) \cdot \left( 1+ \tfrac{3}{160} \, p^2 - \tfrac{1}{960} \, p^3 + \dots \right) \, .
\label{eq_fit_CP_avg}
\end{equation}
After having calculated this average membrane concentration, we can evaluate the average mass transfer coefficient, $\left\langle k \right\rangle$, in analogy with Eq.~\eqref{eq_exponential_law}
\begin{equation}
 \left\langle c\s{m} \right\rangle  = c\s{f} \cdot \exp \left( {J } / {\left\langle k \right\rangle} \right)  \; \leftrightarrow \;   \left\langle k \right\rangle = J / \ln \left( \left\langle c\s{m} \right\rangle /  c\s{f}  \right) \, .
\label{eq_k_average}
\end{equation}
For $X \! = \! 1$, we have $p \! = \! 1.5$, and then the term $\left(1+\tfrac{3}{160}\,p^2 \dots \right)$ has the value 1.04, thus very close to 1.0. Consequently, all terms involving $p^2$ etc., can be neglected and we thus obtain for $X \! < \! 1$ the result that
\begin{equation}
\left\langle k \right\rangle =  \tfrac{4}{3}  \,   {\alpha} \, {J} \, {X^{-1/3}} = \left\langle {\alpha} \right\rangle \, \left(\frac{D^2 \gamma }{L } \right)^{1/3}
\label{eq_k_average_2}
\end{equation}
where now $\left\langle \alpha \right\rangle = \tfrac{4}{3} \cdot 0.662 = 0.883$. The criterion $q \! = \! 1.5$ corresponds to a CP-ratio, here defined as $\left\langle c\s{m} \right\rangle /c\s{f}$, of $\sim \! 3.2$, which is a very high number. Thus, Eq.~\eqref{eq_k_average_2} holds for all realistic situations, i.e., it will be generally valid. Thus we argue that the above two equations, Eq.~\eqref{eq_k_average} and~\eqref{eq_k_average_2}, describe the CP-layer accurately (Eq.~\eqref{eq_k_average} can always be extended to the exponential law that includes solute leakage). These two equations show that the average mass transfer coefficient is not a function of solvent membrane flux, $J$. In the next box we compare with an analysis by Geraldes and Afonso in 2006~\cite{Geraldes_2006}, who mathematically came to the same result but used very different equations. 

\begin{figure}
\centering
\includegraphics[width=0.48\textwidth]{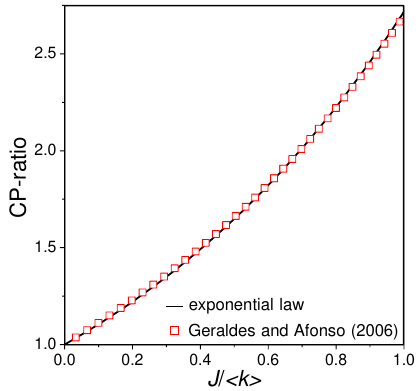}
\vspace{-0pt}
\caption{Comparison of CP model by Geraldes and Afonso (2006) (red squares) and the exponential law (solid line). The two models overlap very closely in the range considered, though the derivation is very different, see the adjoining box. Presented are results of a calculation for the average CP-ratio for a fixed solvent flux, \textit{J}, and for 100\% solute rejection.}
\label{fig_GA}
\end{figure}

\begin{framed}
\noindent \underline{Expressions for CP-layer by Geraldes and Afonso (2006).} Geraldes and Afonso study the same problem as in the present work, and as we will show in this box, come to the exact same outcome, but formulated in a very different manner. They relate concentrations and fluxes at the membrane according to (see their Eq.~(7))
\begin{equation}
1-\frac{c\s{f}}{\left\langle c\s{m} \right\rangle }= \frac{J}{\left\langle k^\textsc{ga} \right\rangle }
\label{eq_Ger_1}
\end{equation}
where we neglect a term $c\s{p}$ which arises via the dead-end equation; thus we assume 100\% rejection for now. Feed concentration $c\s{f}$ can be replaced by $c\s{b}$ or $c_\infty$. Note the crucial absence of any exponential operation. Eq.~\eqref{eq_Ger_1} is arrived at because an equation for the mass transfer layer (which has contributions from convection and diffusion) is solved while setting the convection-term to a constant, equal to the value at the membrane surface. Thus, there is no integration over \textit{x} involved in the derivation of their result.

Their expression for the average mass transfer coefficient, $\left\langle k^\textsc{ga} \right\rangle$, depends on $\kd$ given by Eq.~\eqref{eq__avg_LG}, multiplied by a factor $\Xi$, and that factor depends on a factor $\phi$, which is the solvent membrane flux, \textit{J}, divided by $\kd$. That correlation is $\Xi = \phi + \left(1 + 0.26 \phi^{1.4} \right)^{-1.7}$. 

We can rewrite Eq.~\eqref{eq_Ger_1} to a CP-ratio, $\left\langle c\s{m} \right\rangle / c\s{f}$, and we then obtain
\begin{equation}
\text{CP}^\textsc{ga} = \frac{1}{1-J/\left\langle k^\textsc{ga} \right\rangle} = \frac{\left\langle k^\textsc{ga} \right\rangle}{\left\langle k^\textsc{ga} \right\rangle-J}  = \frac{\Xi}{\Xi-\phi} 
\label{eq_Ger_2}
\end{equation}
and then we implement $\left\langle k^\textsc{ga} \right\rangle = \kd \, \Xi$, and then the function for $\Xi$ as function of $\phi$, with $\phi$ given by $\phi = J / \kd$. The result is shown as squares in Fig.~\ref{fig_GA}. The solid line in Fig.~\ref{fig_GA} is obtained from Eqs.~\eqref{eq_k_average} and~\eqref{eq_k_average_2} above, and as can be observed, the two approaches match very closely. Thus, in a practical sense the two methods support one another because they result in almost exactly the same outcome. 

However, Geraldes and Afonso, because of the chosen framework to define their $\left\langle k^\textsc{ga}\right\rangle$, namely via their Eq.~(7), which leads to the above equations~\eqref{eq_Ger_1} and~\eqref{eq_Ger_2}, do conclude that solvent flux, \textit{J}, enhances the mass transfer coefficient. However, when we analyse the same problem in the context of the exponential law, Eq.~\eqref{eq_k_average}, we come to the opposite conclusion, which is that \textit{J} has no effect on $\left\langle k \right\rangle$. In addition, we would argue that Eqs.~\eqref{eq_k_average} and~\eqref{eq_k_average_2} are easier to analyse than the protocol proposed by Geraldes and Afonso.

\end{framed}

We can use Eq.~\eqref{eq_k_average_2} to calculate $\left\langle k \right\rangle$ for the test cell of Wang \textit{et al}.~\cite{Wang_2021} that was used to study desalination of an NaCl solution by RO. Dimensions of this cell are $H \! = \! 3.0$~mm and $L \! = \! 7.7$~cm, and the crossflow velocity is $U \! = \! 21$~cm/s. For NaCl with a diffusion coefficient of $1.6 \cdot 10^{-9}$~m\textsuperscript{2}/s, we then arrive for $J \! = \! 50$~LMH at $X\! \sim \! 0.2$. If we use the full numerical procedure of Eq.~\eqref{eq_cm_average} we obtain $\left\langle k \right\rangle = 75.0$~LMH, and if we use Eq.~\eqref{eq_k_average_2} we have $\left\langle k \right\rangle = 76.6$~LMH. These values are very close to what was independently measured by Wang \textit{et al}.~\cite{Wang_2021} by measuring water flux with and without added salt (74.9~LMH).

We can also evaluate $\left\langle  k \right\rangle $ for the two types of cells tested in ref.~\cite{Biesheuvel_2023}. For both cells we have a channel height of $H \! = \!  0.7$~mm. For the larger cell design, length is $L \! = \! 19.7$~cm and crossflow velocity is $U \! = \! 21.5$~cm/s, thus $\gamma \! = \! 1840$~s\textsuperscript{-1}. For this design we arrive at $\left\langle k \right\rangle \! = \! 90$~LMH based on Eq.~\eqref{eq_cm_average} (at $J \! = \! 50$~LMH), and at $\left\langle k \right\rangle \! = \!  92$~LMH using Eq.~\eqref{eq_k_average_2}. For the smaller cell ($L \! = \! 7.9$~cm) where crossflow velocity was very high ($U \! = \! 67.5$~cm/s), we have $\gamma \! = \! 5800$~s\textsuperscript{-1}, and $\left\langle k \right\rangle \! = \! 180$~LMH based on Eq.~\eqref{eq_cm_average} ($J=50$~LMH) and $\left\langle k \right\rangle=182$~LMH when using Eq.~\eqref{eq_k_average_2}. These numbers seem very reasonable. 

In the analysis above, we described the situation that in the test cell the solvent membrane flux, \textit{J}, is the same everywhere. But in reality \textit{J} decreases with \textit{z}. How that can be implemented is discussed in section~\ref{section_1D}. There we arrive at the result that for typical test cell conditions, we can also use the equations of the present section, for the 0D model, but with \textit{J} replaced by the average (measured) permeate flux, $\left\langle J \right\rangle$. Thus we can use the exponential law, with all properties averaged, from rewriting Eq.~\eqref{eq_k_average} to
\begin{equation}
 \left\langle c\s{m} \right\rangle  = c\s{f} \cdot \exp \left( {\left\langle J \right\rangle} / {\left\langle k \right\rangle} \right) 
\label{eq_k_average_3}
\end{equation}
which is the exponential for a membrane with 100\% solute rejection. The extension to the exponential law for less than 100\% rejection is then Eq.~(2) in Part~I, with $c\s{m}$, $J$, and $k$ replaced by their averaged (i.e., measured) values, see also Eq.~\eqref{eq_k_average_4} in the next section.\footnote{In solving this 0D model, there is now one inconsistency, namely that the permeate concentration required in this calculation, $c\s{p}$, which can only be calculated from $c\s{p} = \left\langle J\s{s} \right\rangle / \left\langle J \right\rangle $, is not an average permeate concentration, but is the permeate concentration in the exit (effluent). We nevertheless will have to use this equation in the 0D model. In the 1D model, this error is avoided.} 

%\begin{framed}
%\noindent \underline{Higher order averaging of membrane concentration}. As discussed, the above results were based on an averaging process that was relevant when membrane transport is linearly related to $c\s{m}$. However, for a different dependency of membrane solute flux on $c\s{m}$, we need to calculate a different average of the membrane concentration. For instance, when solute flux depends on $c\s{m}$ to the power two, we must evaluate
%
%\begin{equation}
%{\left\langle c_\text{m}^2 \right\rangle} = \frac{1}{L} \int_0^L {c_\text{m}^2\left(z\right)} \, \text{d}z  = {c_\text{f}^2} \cdot {X}^{-1} \cdot \int_0^{X}  \exp\left\{ 2 \, \alpha^{-1} \, \xi^{1/3} \right\} \text{d} \xi 
%\label{eq_cm_squared}
%\end{equation}
%
%\hl{from which follows}
%
%\begin{equation}
%{\left\langle c_\text{m}^2 \right\rangle}  \sim {c_\text{f}^2} \cdot \exp\left( \tfrac{3}{2} \cdot \alpha^{-1} \cdot {X^{1/3}} \right)  \cdot f\left(X\right)
%\label{eq_cm_squared_2}
%\end{equation}
%
%where $f\left(X\right)$ is given by $f\left(X\right)=1+3/40\cdot q^2 - 1/120 \cdot q^3 + \dots$. For $J \! = \! 50$~LMH and $X \! = \! 0.2$, Eq.~\eqref{eq_cm_squared_2} is only 5\% below the prediction of Eq.~\eqref{eq_cm_squared}. The alternative to implementing an equation such as Eq.~\eqref{eq_cm_squared_2}, is to assume that the regular averaging of $c\s{m}$, as discussed before this box, is sufficiently accurate (ideally checked by comparison with 1D or 2D models), or, not to average at all, and use the 1D and 2D models of sections~\ref{section_1D} and~\ref{section_2D}. 
%\end{framed}

\section{One-dimensional model for reverse osmosis including concentration polarization}
\label{section_1D}

In the previous section a method was discussed where the entire module or test cell is described by a small number of equations, i.e., a `zero-dimensional' (0D) approach is used. More detailed is a one-dimensional (1D) model where a position-dependent expression for \textit{k} is implemented, and we track precisely how the membrane concentration, $c\s{m}$, increases in \textit{z}-direction. Because of the increase in $c\s{m}$, also the osmotic pressure at the membrane surface increases, and thus the solvent flux through the membrane, \textit{J}, will decrease. In the 1D model all of these changes through the channel are incorporated. We do not average over the membrane surface, and thus we avoid several problematic issues discussed in the previous section that relate to the process of averaging. 

What we propose is to use Eq.~\eqref{eq_k_definition} in the 1D model, combined with the exponential law (with or without solute leakage), to relate the local membrane concentration, $c\s{m}\left(z\right)$, to the local permeate (water) flux $J\left(z\right)$. In this equation also the concentration outside the CP layer is required, for which we use the average, or `mixed cup' concentration, which in such 1D models is simply called concentration, \textit{c}, without an extra index, see ref.~\cite{Biesheuvel_2022}, and we follow that convention.

In the \textit{z}-direction along the membrane we assume all transport is by convection, thus neglecting diffusion. A solute mass balance is then (see Eq.~(38) in ref.~\cite{Biesheuvel_2022})
\begin{equation}
\frac{\partial }{\partial {z}} \left( \phi\s{v} \cdot c  \right) = - W J\s{s}
\label{eq_1D_mass_balance_gen}
\end{equation} 
where $\phi\s{v}$ is the volumetric flow rate in the feed channel, and $J\s{s}$ is the membrane solute (salt) flux. When we implement that $\phi\s{v}=U\cdot W\cdot H$ and $A=W\cdot L$, we can rewrite Eq.~\eqref{eq_1D_mass_balance_gen} to
\begin{equation}
\frac{\partial }{\partial {z}} \left( \vphantom{\phi\s{v}\cdot c} U \cdot c \right) = - \frac{J\s{s}}{H} \, .
\label{eq_1D_mass_balance_2}
\end{equation} 
A balance for the total volume flow in the feed channel is~\cite{Junker_2021}
\begin{equation}
\frac{\partial U }{\partial {z}} = - \frac{J}{H}  \, .
\end{equation} 

In the 1D model, we describe transport across the CP layer by the exponential law for non-perfect rejection (see Eq.~(9) in ref.~\cite{Biesheuvel_2023})
\begin{equation}
c\s{m} = \left( c\s{f} - \frac{J\s{s}}{ J } \right) \cdot \exp \left( \frac{ J }{k } \right) + \frac{J\s{s} }{J }
\label{eq_k_average_4}
\end{equation}
with all parameters in Eq.~\eqref{eq_k_average_4} dependent on \textit{z}-coordinate. We do not use a fixed value for \textit{k}, but use the result of Eq.~\eqref{eq_k_definition} ($\alpha = 0.662$, $\gamma=6 \, U / H$), such that \textit{k} will change from infinity right at the start of the channel, to lower and lower values the further down we go into the channel.\footnote{In the calculation of \textit{k}, it makes hardly any difference if the decrease in \textit{U} through the channel is made to change $\gamma$, or if $\gamma$ is set to the value at the inlet.} 

For the flux of salt across the membrane, $J\s{s}$, we can use the solution-friction (SF) model, which provides accurate equations that include the effect of membrane charge, see for example Eq.~(8) in ref.~\cite{Biesheuvel_2023}. For an uncharged membrane (and for neutral solutes), for desalination with RO, this simplifies to the classical expression
\begin{equation}
J\s{s} = B \cdot \left(c\s{m} - c\s{p} \right)
\label{eq_B_solute}
\end{equation} 
which will be used in the present calculations. 

The flux of solvent across the membrane, \textit{J}, is given by
\begin{equation}
J = A \cdot \left( \Delta P^{\text{h},\infty} - \sigma_i \cdot \Delta \Pi^\infty\right)
\label{eq_J_A}
\end{equation}
where \textit{A} is the membrane water permeability, which can be expressed in in LMH/bar. The reflection coefficient, $\sigma_i$, is a constant that depends on the properties of the membrane and solute~\cite{Biesheuvel_2022}. We will assume that in the entire cell we have the same hydrostatic pressure difference across the membrane, $\Delta P^{\text{h},\infty}$, i.e., we neglect the axial gradient in pressure (which is minor in a short test cell). Assuming ideal solutions, the osmotic pressure difference across the membrane is the concentration difference $c\s{m}-c\s{p}$, multiplied by \textit{RT}, and for a salt multiplied by an extra factor (for instance, for a 1:1 salt this factor is 2). Below we evaluate the osmotic pressure more precisely.

Finally, we need to calculate the permeate concentration $c\s{p}$, which can be done in one of several ways. When there is no flow of permeate along the membrane, and no mixing, the dead-end equation applies at each position, $c\s{p}=J\s{s}/J$. If there is a flow channel on the permeate side, with water flowing in the same direction as on the feed side, then $c\s{p}$ is calculated from an overall solute balance
\begin{equation}
U_0 \cdot c_0 = U \left(z\right) \cdot c\left(z\right) + \left(U_0 - U\left(z\right) \right) \cdot c\s{p}\left(z\right)
\label{eq_overall_module_balance_cp}
\end{equation}
where variables with index 0 refer to values at the entrance of the channel on the feed side.\footnote{Eq.~\eqref{eq_overall_module_balance_cp} follows from $H U_0 c_0 = H U\left(z\right)c\left(z\right) + H\s{p} U\s{p}\left(z\right) c\s{p}\left(z\right)$, where $H\s{p}$ is the height of the permeate channel and $U\s{p}$ the crossflow velocity there. Conservation of total volume flow is given by $H U_0 = H U\left(z\right) + H\s{p} U\s{p}\left(z\right) $ and combination of these two equations results in Eq.~\eqref{eq_overall_module_balance_cp}.} A third option is to set the permeate concentration to zero. Here we use the second approach, using an overall solute balance. 

\begin{figure}[!ht]
\centering
\includegraphics[width=0.75\textwidth]{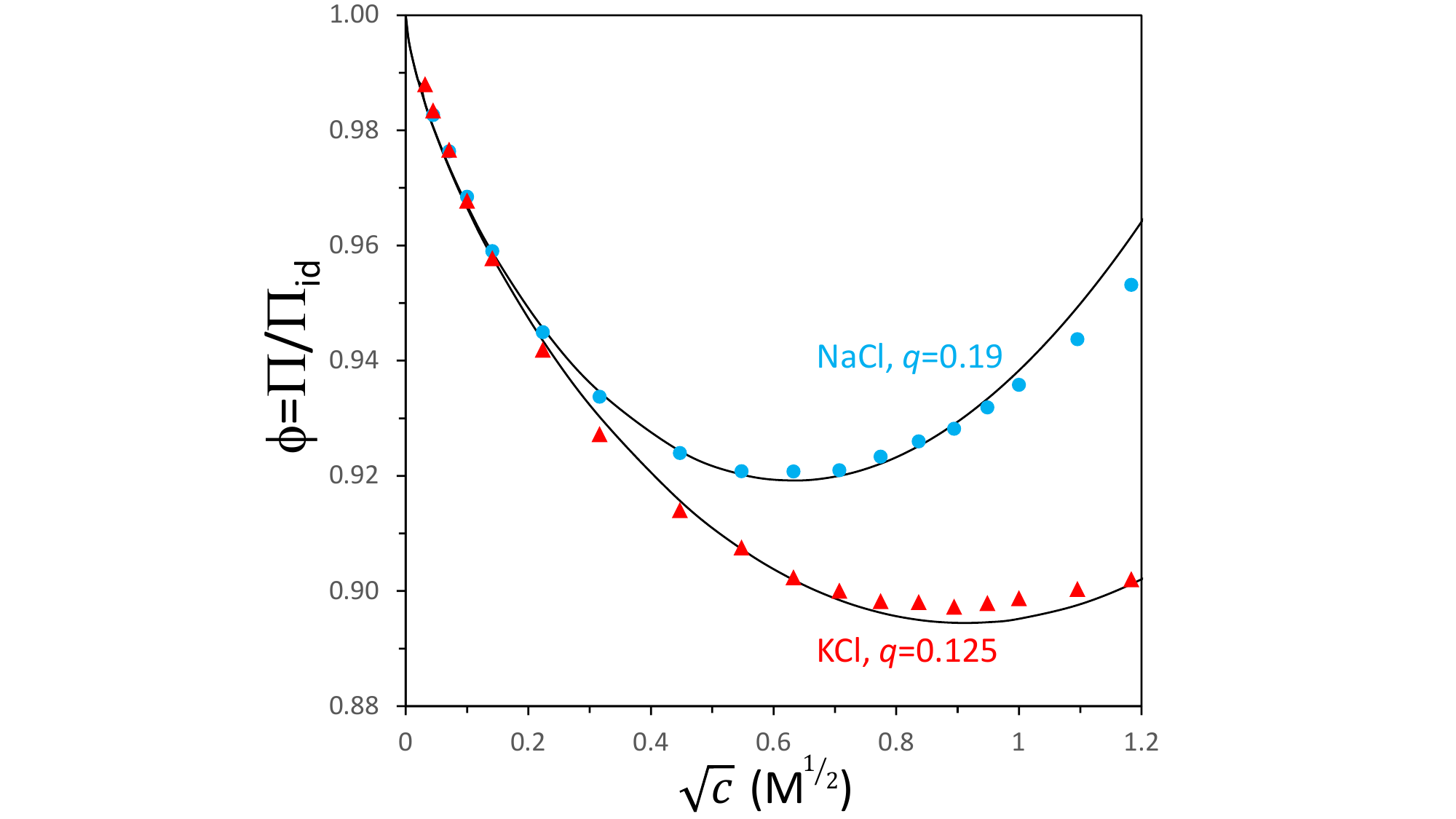}
\vspace{-0pt}
\caption{The osmotic coefficient, $\phi$, of a solution of NaCl or KCl as function of salt concentration \textit{c}, where $\phi = \Pi / \Pi\s{id}$ with $\Pi$ osmotic pressure and $\Pi\s{id}=2 c RT$ the ideal osmotic pressure. Lines according to Eq.~\eqref{eq_Pi_nonideal} and points are data.}
\label{fig_osm_coeff}
\end{figure}

The advantage of the 1D model is that it not only applies to neutral solutes, but can also be used for 1:1 salt solutions (or other binary salt solutions), using any type of equation for salt flux across the membrane, such as for instance Eq.~(8) in ref.~\cite{Biesheuvel_2023}, that has a non-linear dependence on $c\s{m}$. We can also include an activity correction to the osmotic pressure, which makes the expression for solvent flux non-linear with respect to $c\s{m}$ and $c\s{p}$. We include this effect in all our calculations (for the 0D, 1D, and 2D models), using an improvement of Eq.~(16) from ref.~\cite{Biesheuvel_2023}, which is
\begin{equation}
\phi = \frac{\Pi}{\Pi\s{id}} =1 - \gamma \cdot {\overline{c}}^{1/3} - 8/5 \cdot \gamma^2 \cdot {\overline{c}}^{2/3} + 192 \cdot \gamma^3 \cdot q \cdot \overline{c}  
\label{eq_Pi_nonideal}
\end{equation}
where $\phi$ is the osmotic coefficient and for a 1:1 salt the ideal osmotic pressure is $\Pi\s{id} = 2 c RT$. In Eq.~\eqref{eq_Pi_nonideal}, $\overline{c}=c/c\s{ref}$, with $c\s{ref}=1$~mM, and $\gamma$ is a factor given by $\gamma=0.0151$. For NaCl the sizefactor \textit{q} is  $q \! = \! 0.19$, and for KCl it is $q \! = \! 0.125$. We plot this function against the square root of concentration in Fig.~\ref{fig_osm_coeff} both for NaCl and KCl solutions, and compare with data~\cite{Hamer_Wu_1972}. We notice that this modified expression is quite accurate up to a salt concentration of $c_\infty \sim 1.5$~M. % these data for \phi in Hamer and Wu, corresponding to Handbook Chemistry Physics. interestingly, these data follow exactly from a certain type of integration of data for ln\gamma. This makes sense because measuring osmotic coefficients directly, is hardly possible.
In the calculations in the present work we use Eq.~\eqref{eq_J_A}, where we set the reflection coefficient to zero, and implement that $\Delta\Pi^\infty = \Pi\s{m} -\Pi\s{p}$. We evaluate these two osmotic pressures using Eq.~\eqref{eq_Pi_nonideal} for an NaCl solution ($q \! = \! 0.19$). 

\begin{figure}
\centering
\includegraphics[width=0.75\textwidth]{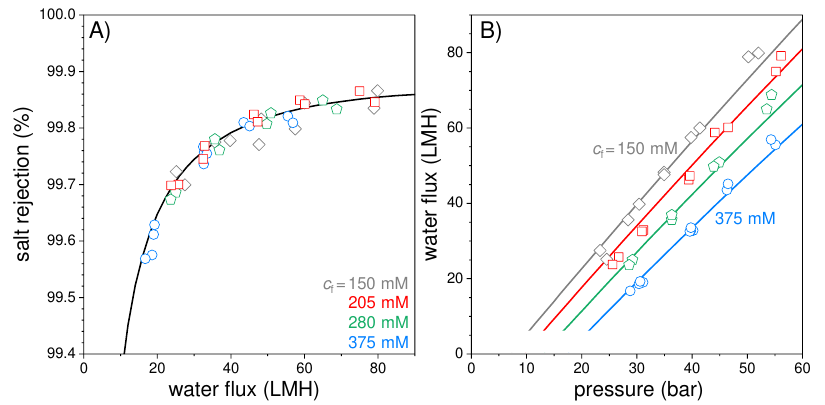}
\vspace{-0pt}
\caption{Data and calculations of an RO experiment using a SW30-HRLE membrane~\cite{Blankert_2022, Biesheuvel_2023}. Calculation results are based on a zero-dimensional (0D) and one-dimensional (1D) model. The theoretical lines overlap almost completely so only one set of lines is plotted.}
\label{fig_0D_1D}
\end{figure}

In the 1D model discussed here, and likewise for the 2D model discussed next, these modifications to the expressions for flow of solvent and salt across the membrane can be easily implemented, without conflicting with the modelling framework. This is different for the 0D approach, where the standard way of averaging to arrive at $\left\langle c\s{m} \right\rangle$ assumes that all membrane fluxes are proportional to $c\s{m}$. When they are not, a more complicated averaging procedure is necessary, but it may then be preferable to analyse a 1D model instead.

We compare the 0D and 1D model for a typical test experiment for the characterization of an RO membrane. To that end we use data obtained with the test cell from ref.~\cite{Blankert_2022} ($H \! = \! 0.711$~mm, $L \! = \! 20$~cm, $U \! = \! 20$~cm/s) for the membrane SW30-HRLE, see Table~S1 in ref.~\cite{Biesheuvel_2023}. When we use model settings that describe the data, the predictions of the 0D and 1D models are virtually identical and thus we only plot one set of lines (membrane properties are $A=1.95$~LMH/bar and $B=0.060$~LMH). We will later on show that the 1D model closely matches the 2D model, and thus we can conclude that (at least in this case) the 0D model can be reliably used to describe membrane performance in a typical test cell geometry. For these test cell conditions, for an  NaCl solution ($D\s{hm,NaCl}=1.6\cdot 10^{-9}$~m\textsuperscript{2}/s), and for an empty channel, we calculate that $\left\langle k \right\rangle \! \sim \! 90$ LMH. However, with this value of $\left\langle k \right\rangle$, the fit to data is not optimal (not reported), which must be because in this experiment the feed channel has a spacer material. A spacer material enhances mass transfer because of higher shear rates near the membrane. To obtain a good fit, we use a higher transfer coefficient of $\left\langle k \right\rangle \! = \! 120$~LMH, resulting in the curves plotted in Fig.~\ref{fig_0D_1D}, where we show a very good fit between theory and data for salt and water flux as function of pressure. 

\begin{figure}[!ht]
\centering
\includegraphics[width=1.0\textwidth]{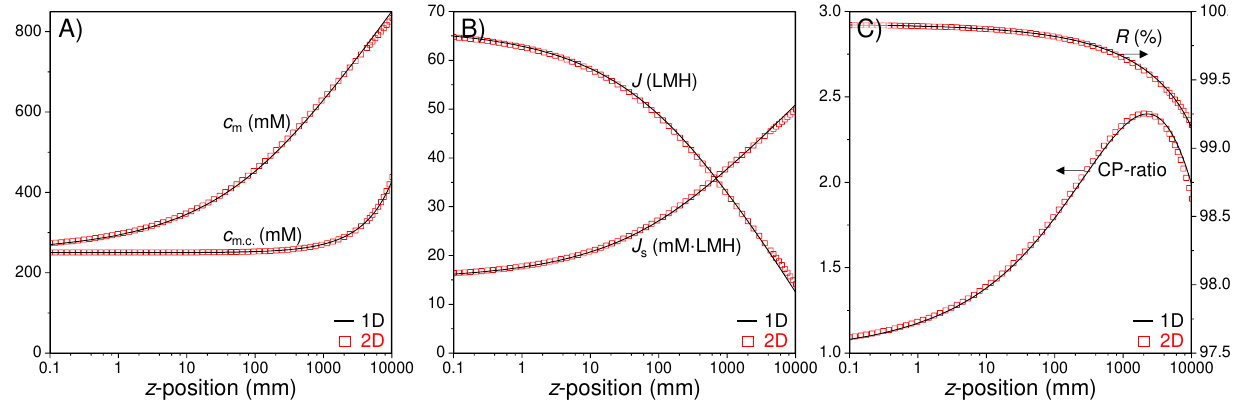}
\vspace{-20pt}
\caption{Calculation results of the 1D model (lines) and 2D model (red squares) for the flow in an RO feed channel including solvent and solute passing the membrane. All parameter settings reported in main text. A) Calculated membrane concentration, $c\s{m}$, and average, mixed cup, concentration, $c\s{m.c.}$; B) Transmembrane flux of water, $J$, and salt, $J\s{s}$; C) Salt rejection, \textit{R}, and CP-ratio, $c\s{m}/c\s{m.c.}$, all as function of axial coordinate \textit{z}.}
\label{fig_1D_2D}
\end{figure}

It is interesting to evaluate the 1D model because it leads to more insight in local concentrations and fluxes in the cell.\footnote{And it useful to evaluate the 1D model with the aim to compare with predictions of the 0D model, to check if the assumptions made in the 0D model are correct.} For instance, the 1D calculation shows how the transmembrane water flux, \textit{J}, changes between the inlet and outlet. For an applied pressure of $\Delta P^{\text{h},\infty} \! = \! 45.7$~bar, \textit{J} changes from 63~LMH initially to 45~LMH at the end of the module which averages out to $\left\langle J \right\rangle = 50$~LMH. The concentration at the membrane, $c\s{m}$, increases from $c\s{m} \! = \! 250$~mM at the inlet (no CP effect at the start) to a value after $L\!=\!20$~cm of $c\s{m} \! = \!  500$~mM, and the average is $\left\langle c\s{m} \right\rangle \! = \!  440$~mM.\footnote{The initial increase is extremely rapid. Only 100~$\mu$m into the channel, we already have $c\s{m}=270$~mM instead of the inlet value of 250~mM.}\textsuperscript{,}\footnote{In this calculation, as well as those reported in Figs.~\ref{fig_1D_2D} and~\ref{fig_CP_layer}, we use the input numbers as given above for an open channel, without including a correction for the presence of spacer material.} The average concentration in the channel, \textit{c}, only increases slightly, from 250 to 255~mM. 

Finally, we compare these calculations of the 1D model with the full 2D model that we explain in detail in section~\ref{section_2D}. In Fig.~\ref{fig_1D_2D} we compare these two models with respect to the mixed cup concentration, $c\s{m.c.}$, membrane concentration, $c\s{m}$, membrane flux of solvent and solutes, \textit{J} and $J\s{s}$, and solute rejection, \textit{R}, and CP-ratio ($c\s{m}/c\s{m.c.}$), plotted versus position in the test cell, \textit{z}. We evaluate here channels that are much longer than the original test cell that had a length of $L \! = \! 20$~cm.\footnote{Because in the channel there is no axial diffusion, at any particular \textit{z}-value, $z^*$, the presented results for $0<z<z^*$ are indicative of a module of length $z^*$, and results for $z > z^*$ can be neglected.} The first observation is the very close agreement of the outputs of these two models, with only a small difference developing beyond $z \! = \! 1$~m. Both models predict that $c\s{m}$ and $c\s{m.c.}$ both increase, with $c\s{m}$ increasing faster. In both models the CP-ratio reaches a maximum of $\text{CP}\s{max} \! \sim \!  2.4$, and then decreases again. With increasing \textit{z}, the membrane solvent flux decreases, while membrane salt flux goes up, and thus salt rejection steadily decreases with \textit{z}. The main conclusion of the close match between these two models is that the 1D model is accurate, so the expression for $k\left(z\right)$ given by Eq.~\eqref{eq_k_definition} can also be used when the boundary solvent flux, \textit{J}, is not constant. 

\section{Numerical two-dimensional model for concentration polarization} 
\label{section_2D}

In this last section we discuss the procedure to set up a complete 2D calculation. This was already done by Sherwood \textit{et al}.~\cite{Sherwood_1965} for the case of a membrane placed on both sides of the channel, for a constant solvent flux \textit{J}, and for 100\% solute rejection. In our calculations that we report in Fig.~\ref{fig_1D_2D} this model is modified to describe a channel with only one membrane, and we include that the membrane flux of solvent decreases and of solute increases, by implementing Eqs.~\eqref{eq_B_solute}--\eqref{eq_overall_module_balance_cp}.

In the 2D model it is assumed that the fluid velocity profile in \textit{z}-direction is always parabolic, with a no-slip boundary condition on the two sides of the channel. So we assume that the parabolic profile is already established right at the point that the solution `hits' the membrane. While solvent flows sideways out of the cell, the average velocity, \textit{U}, decreases, but the velocity profile remains parabolic. Because of that constraint, we know exactly how the \textit{x}-component of the solvent velocity changes with \textit{x}, where \textit{x} is the coordinate at right angles to the \textit{z}-direction of fluid flow, i.e., towards the membrane. 
At the membrane, this \textit{x}-component of the fluid velocity has the value $J\left(z\right)$, i.e., the solvent flux through the membrane~\cite{Bouchard_1994}. Interestingly, next to the membrane, $v_x$ is fairly constant for the first 10\% of the channel height \textit{H}, and only beyond this region, $v_x$ starts to decrease to reach $v_x\!=\!0$ at some point. In case of one membrane, this point is at the other side of the channel, but with a membrane on each side of the channel, $v_x \! = \! 0$ is reached at the center plane of the channel. For a single membrane, the full expression for $v_x$ is Eq.~(11) in ref.~\cite{Johnston_2023}, which is
\begin{equation}
\frac{v_x}{J} =  1-3 \left(\frac{\widetilde{x}}{H} \right)^2 + 2 \left(\frac{\widetilde{x}}{H}\right)^3 =  
 3 \left(\frac{x\vphantom{^*}}{H}\right)^2 - 2 \left(\frac{x\vphantom{^*}}{H}\right)^3 
\label{eq_vx_one_membrane}
\end{equation} 
where $\widetilde{x}$ is a coordinate that starts at the membrane and points into solution, while \textit{x} starts at the backside and points towards the membrane ($x+\widetilde{x}=H$). Eq.~\eqref{eq_vx_one_membrane} shows that halfway in the channel, $v_x$ is half of \textit{J}.
For the case of two membranes, instead of Eq.~\eqref{eq_vx_one_membrane} we have~\cite{Brian_1965,Kim_2005}
\begin{equation}
\frac{v_x }{ J } =  1- 6 \left(\frac{\widetilde{x}}{H} \right)^2 + 4 \left(\frac{\widetilde{x}}{H}\right)^3  = \frac{3}{2}  \left(\frac{x^*}{h}\right) - \frac{1}{2}\left(\frac{x^*}{h}\right)^3 
\label{eq_vx_two_membrane}
\end{equation} 
where \textit{h} is the channel half height ($h=H/2$), and $x^*$ is a coordinate axis starting at the center plane. Note that all velocities above are dependent on both \textit{x} and \textit{z}.

Besides one of these two expressions for $v_x$, we have the differential mass balance
\begin{equation}
\frac{\partial c}{\partial t} = - \frac{\partial }{\partial z} \left( v_z c \right) - \frac{\partial }{\partial x} \left(  v_x c \right) + D \frac{\partial^2 c}{\partial x^2}
\end{equation}
which for steady state ($\partial c / \partial t = 0$) and because of continuity of volume, $\partial v_z / \partial z + \partial v_x / \partial x = 0$, simplifies to
\begin{equation}
v_z \frac{\partial c}{\partial z} + v_x \frac{\partial c}{\partial x} = D \frac{\partial^2 c}{\partial x^2} \, .
\end{equation}
For a single membrane, then on the backside of the channel we have zero flux into the wall for solvent and solute, and thus we have $\partial c / \partial x = 0$  there. In case of two membranes, we have this symmetry condition at the center plane. At the membrane we have a certain solvent flux calculated from $c\s{m}$ and the membrane equations. This $c\s{m}$ also leads to a prediction for $J\s{s}$ and thus to $c\s{p}$. We can evaluate a flux for solutes at the channel/membrane boundary according to $J\s{s}=J c\s{m} - D \left. \partial c / \partial x\right|\s{m}$, or we evaluate an overall mass balance, expressing that the change with \textit{z} of the total molar flow rate of solutes (average crossflow velocity times mixed cup concentration) divided by channel height \textit{H} for one membrane, or divided by half-thickness \textit{h} for two membranes, equals the membrane solute flux, $J\s{s}$~\cite{Biesheuvel_2023}, see Eq.~\eqref{eq_1D_mass_balance_2} in case of one membrane (with \textit{c} there replaced by $c\s{m.c.}$). We use the overall mass balance approach but for 100 or so gridpoints in \textit{x}-direction, the difference between that method and numerically evaluating $J\s{s}=J c\s{m} - D \left. \partial c / \partial x\right|\s{m}$, is small.

The 2D model is easily solved after discretization in \textit{z}-direction according to the implicit Euler-scheme, and in the \textit{x}-direction according to a central difference scheme, which means that in the \textit{x}-direction, all nodes (gridpoints) in the calculation are solved simultaneously. Results of the 2D model were reported in Fig.~\ref{fig_cmcf} for the symmetric Sherwood \textit{et al}.~ geometry with two membranes and a fixed solvent membrane flux, \textit{J}, and in Fig.~\ref{fig_1D_2D} for a channel with a single membrane, a non-constant \textit{J} and non-zero solute flux, $J\s{s}$. We finally evaluate the 2D model once again in a calculation similar to the one used for Fig.~\ref{fig_cmcf}, and now plot the function $\Gamma = c\s{m} / c\s{m.c.} - 1$ against $\xi$, exactly as reported by Sherwood \textit{et al}.~(1965)~\cite{Sherwood_1965}. The symbols are calculations by Sherwood \textit{et al}., performed at an MIT computation center at the time. We make the same calculation and we find the exact same output. This precise correspondence is very heartening. Interestingly, our calculations show that right after each of the last calculation points reported by Sherwood \textit{et al}. (right after the last circle, square, and triangle), the calculation comes to a halt, because all the water has been extracted from the feed channel, i.e., water recovery goes to 100\%. In Sherwood \textit{et al}., there is no mention of the impossibility of  calculations beyond the last reported point, and it is not sure if the authors realized that. This point seems also to have been missed by Probstein (1989)~\cite{Probstein_1989} who even introduces plateau regions for $\Gamma$ that extend to $\xi \rightarrow \infty$, but those plateaus are misleading: either, for a fixed \textit{J} the calculation ends because there is no water left in the feed channel, %we reach 100\% water recovery, 
or when \textit{J} is allowed to decrease with \textit{z}, then at a much earlier point the $\Gamma$-function decreases to zero because the lowering of \textit{J} results in $c\s{m}$ and $c\s{m.c.}$ to converge. 
For a fixed \textit{J} the maximum in $\xi$ can be calculated as $\xi\s{max} = 1/ \left(3 a^2\right)$, which for $a=0.50$ results in $\xi\s{max}=1.33$, which exactly agrees with the point of steep decline in Fig.~\ref{fig_1D_2D} for $a \! = \! 0.50$ (and this criterion also agrees with the points of steep decline for the other \textit{a}-values evaluated). This calculation is extended by Brian~\cite{Brian_1965} who uses an expression for the solvent membrane flux that depends on the osmotic pressure (he uses Eq.~\eqref{eq_J_A} with $\Pi \! = \! \Pi\s{id}$). % Brian includes in the osmotic pressure difference also that of the permeate, which results in his osmotic term to be \Pi\s{m} times R. and R is intrinsic, i.e., based on 1-cp/cm
In this calculation, $\Gamma$ has a maximum at a certain $\xi$ and then decreases again. He reports a maximum CP of $\sim \!1.8$ (see his Figs.~2 and~5, $\beta=1$). Other results he presents are for an intrinsic rejection, \textit{R}, defined as $R=1-c\s{p}/c\s{m}$, that is less than 100\%, implementing the dead-end equation as well.\footnote{However, a fixed intrinsic rejection is only possible when the solute flux is described by $J\s{s}=c\s{m} \cdot J \cdot \left(1-R\right)$, which is different from the classical result for neutral molecules that $J\s{s}=B \cdot \left(c\s{m} - c\s{p} \right)$, see also ref.~\cite{Kim_2005}.}

\begin{figure}[!ht]
\centering
\includegraphics[width=0.6\textwidth]{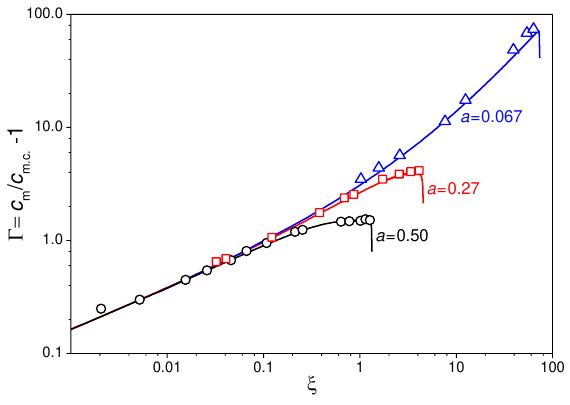}
\vspace{-10pt}
\caption{Membrane concentration, $c\s{m}$, divided by mixed cup concentration, $c\s{m.c.}$, as function of dimensionless position, $\xi$, in RO feed channel. Full 2D calculation for constant permeate flux $J$, with membranes on both sides of the channel. Lines are based on our own numerical calculations, while symbols are for the same calculation by Sherwood \textit{et al}.~(1965).} 
\label{fig_sherwood}
\end{figure}

\begin{figure}
\centering
\includegraphics[width=0.93\textwidth]{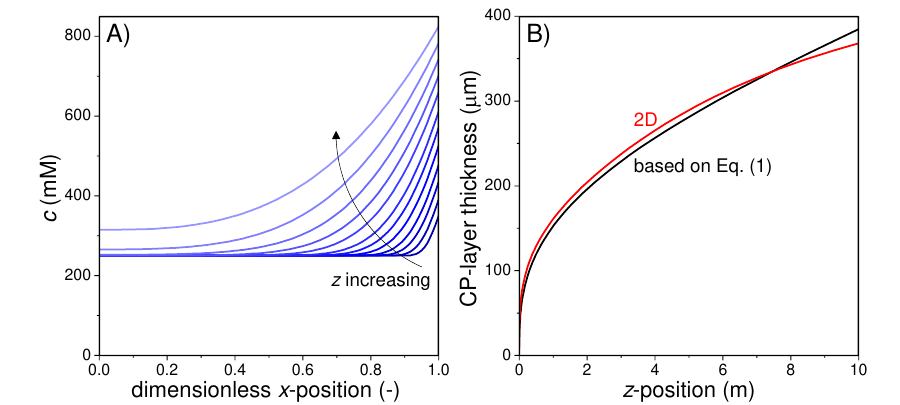}
\vspace{-0pt}
\caption{Structure of the concentration polarization (CP) layer in a rectangular channel with a membrane on one side (on the right in panel A) as function of axial coordinate \textit{z} based on the 2D model (parameters as in Fig.~\ref{fig_1D_2D}). A) Profiles of salt concentration across the thickness of the channel ($H \! = \! 0.71$~mm), % 711 micron
at distinct \textit{z}-coordinates, going downstream in direction of arrow. B) Thickness of CP layer derived from 2D model and from analytical equation for \textit{k}, Eq.~\eqref{eq_k_definition}.}
\label{fig_CP_layer}
\end{figure}

Finally, we discuss the thickness and structure of the CP-layer, by analysing profiles based on a calculation with a single membrane and decreasing \textit{J}, as reported in Fig.~\ref{fig_1D_2D}. A range of concentration profiles $c\left(x\right)$ for increasing \textit{z}-coordinates is shown in Fig.~\ref{fig_CP_layer}A, and we can observe the approximately exponential shape of each profile. With increasing \textit{z}, concentrations steadily go up everywhere in the channel and the CP-layer thickness increases. We can define a CP layer thickness, $\delta$ as the product of three factors: 1. the inverse of the slope (gradient) of concentration at the membrane (in \textit{x}-direction); 2. the concentration difference between the membrane surface and the left-most position (at that same \textit{z}-coordinate); and 3. an empirical multiplication factor, for which we use 1.5. The resulting thickness $\delta$ is plotted versus position \textit{z} in Fig.~\ref{fig_CP_layer}B, as well as the predicted thickness based on $\delta=k/D$ with \textit{k} calculated from Eq.~\eqref{eq_k_definition} (with $\alpha=0.662$). Though the two curves do not match up perfectly, they show a similar trend. We notice that the CP layer is quite thin for typical dimensions of a test cell. In this case the test cell was $L=20$~cm, and after that distance the CP-layer reached a thickness of $\delta \sim 100~\mu$m, about 1/8th of the channel height. But we do see that for longer channels the CP layer is no longer thin, but now occupies a significant part of the channel, so ideally it is no longer treated as a boundary condition in a mathematical theory, for instance beyond $z \! = \! 1$~m in the calculations of Fig.~\ref{fig_CP_layer}. However, the quite close fit between the 1D and 2D models, as was demonstrated in Fig.~\ref{fig_1D_2D}, indicates that the error is not that large when expressions are used for the CP layer that do treat it mathematically as a boundary condition. 

\section*{Conclusions}

We analysed concentration polarization in pressure driven membrane separation, such as by reverse osmosis and nanofiltration, using zero-dimensional, one-dimensional, and two-dimensional models of an open feed channel. The 2D model makes the least number of assumptions and fully resolves flow and concentration profiles in the feed channel, while in the 1D model a novel expression for the CP layer mass transfer coefficient is implemented that works very well. The 0D model is the most concise but an averaging procedure is required that introduces various assumptions. In calculations for an uncharged membrane, results of the three geometries match closely, and thus assumptions made in the 0D and 1D models were accurate. All of these results are for a solution with a single solute or salt. When we have many types of solutes and ions, the complete 2D model may be the only way to obtain accurate results for concentrations in the feed channel.

A new analytical equation is derived for the mass transfer coefficient in the CP layer, \textit{k}, based on an earlier expression derived by Sherwood \textit{et al}.~in 1965. It looks similar, with only a different prefactor, to a formulation of $k$ based on the standard Lévêque-Graetz problem (that is based on a fixed wall concentration as boundary condition), but now this expression is based on detailed comparison with numerical calculations that use the correct formulation of the boundary condition, which is a relation for solute flux that includes diffusion and convection. Interestingly, the solvent membrane flux, \textit{J}, in practice does not influence the mass transfer coefficient, and thus the correlation for \textit{k} can also be used for a module calculation where \textit{J} decreases along the membrane. 

The problem in the 0D approach is that an averaging is required and the type of averaging depends on the type of membrane transport equation. That problem is resolved when a 1D model formulation is used, which allows for any set of membrane equations without changing the model structure for the feed channel. The 2D model does not assume anything about the CP layer but numerically resolves its entire structure. Such a 2D calculation was already made by Sherwood \textit{et al}.~(1965) and we confirm correctness of their numerical calculation. However, a statement made later by Probstein (1989) of the CP-ratio levelling off beyond a certain position, turns out to be incorrect. Instead, beyond the last calculation result reported by Sherwood \textit{et al}., perhaps unknown to them, their calculation (which is for a fixed solvent flux) will terminate because all the water is gone from the feed channel. A more realistic calculation with a variable \textit{J}, results in the CP-ratio dropping off earlier and more gradually, without the calculation suddenly terminating, as already analysed by Brian, also in 1965.

%\hl{An important conclusion is that when evaluating a series of RO experiments and fitting a transport model to membrane performance data, one should implement calculated values of $k$ or $\left\langle k \right\rangle$, or in any case, the values that are used should correlate in a logical way to cell design and flow conditions. For instance, when there are data in a range of crossflow velocities, \textit{U}, the CP layer mass transfer coefficient \textit{k} should approximately change according to Eq.~\eqref{eq_k_average_2}, thus increasing with crossflow velocity, \textit{U}, to a 1/3 power. And when testing different membranes in the same test cell at the same crossflow velocity, the same \textit{k}-value must be used irrespective of permeate solvent flux, \textit{J}, because \textit{J} has hardly any effect on \textit{k}.}

We can conclude that the CP layer that forms in an open channel can be described in detail at various levels of analysis. Therefore, for membrane characterization in a test cell, we advise to use an open channel, and not place a spacer material in the channel, because for an open channel the characteristics of the CP layer are well established, and the CP effect can be accurately accounted for in the experimental analysis.


\begin{thebibliography}{100}

\bibitem{CP_arxiv_2023}
P.M. Biesheuvel, S. Porada, I. Ryzhkov, and M. Elimelech, ``General validity of the exponential law for the effect of concentration polarization in reverse osmosis in a stirred-cell geometry, including an activity correction for 1:1 salt solutions,'' \textit{Arxiv}:2308.14233 (2023).

\bibitem{Dresner_1964}
L.~Dresner, ``Boundary layer build-up in the demineralization of salt water by reverse osmosis,'' Oak Ridge Natl. Lab., Report No. ORNL-3621 (1964). \textit{https://www.osti.gov/servlets/purl/6365302.} 

%\bibitem{Gill_1965} W.N. Gill, C. Tien, and D.W. Zeh, ``Concentration Polarization Effects in a Reverse Osmosis System,'' \textit{Ind. Eng. Chem. Fund.} \textbf{4}, 433--439 (1965). % not important enough to cite

\bibitem{Brian_1965}
P.L.T. Brian, ``concentration polarization in reverse osmosis desalination with variable flux and incomplete salt rejection,'' \textit{Ind. Eng. Chem. Fund.} \textbf{4}, 439--445 (1965). % geweldige discussie over dit artikel in Bouchard JMS 1994. % https://en.wikipedia.org/wiki/P._L._Thibaut_Brian
% https://news.mit.edu/2018/pl-thibaut-brian-professor-emeritus-chemical-engineering-dies-0418

\bibitem{Oren_2021} Y.S. Oren, V. Freger, and O. Nir, ``New compact expressions for concentration-polarization of trace-ions in pressure-driven membrane processes,'' \textit{J. Membr. Sci.} 100003 (2021).

\bibitem{Rohlfs_2016}
W. Rohlfs, G.P. Thiel, and J.H. Lienhard, ``Modeling reverse osmosis element design using superposition and an analogy to convective heat transfer,'' \textit{J. Membr. Sci.} \textbf{512}, 38--49 (2016).

\bibitem{Sherwood_1965}
T.K. Sherwood, P.L.T. Brian, R.E. Fisher, and L. Dresner, ``Salt concentration at phase boundaries in desalination by reverse osmosis,'' \textit{I\&EC Fundamentals} \textbf{4}, 113--118 (1965).

%Sherwood, T. K.; Brian, P. L. T. & Fisher, R. E. Salt Concentration at Phase Boundaries in Desalination Processes, report, March 1964
%https://digital.library.unt.edu/ark:/67531/metadc11647/m1/2/
%office of saline water research and development progress report no 95
%The same also published as:
%Desalination Research Laboratory, Department of Chemical Engineering,
%Massachusetts Institute of Technology, Rept. 295-1 (1963).

\bibitem{Zydney_1986}
A.L. Zydney and C.K. Colton, ``A concentration polarization model for the filtrate flux in cross-flow microfiltration of particulate suspensions,'' \textit{Chem. Eng. Comm.} \textbf{47}, 1--21 (1986).

\bibitem{De_1997}
S. De, and P.K. Bhattacharya, ``Prediction of mass-transfer coefficient with suction in the applications of reverse osmosis and ultrafiltration,'' \textit{J. Membr. Sci.} \textbf{128}, 119--131 (1997).

\bibitem{Kim_2005}
S. Kim and E.M.V. Hoek, ``Modeling concentration polarization in reverse osmosis
processes,'' \textit{Desalination} \textbf{186}, 111--128 (2005).

\bibitem{Zydney_1997}
A.L. Zydney, ``Stagnant film model for concentration polarization in membrane systems,'' \textit{J.~Membr. Sci.} \textbf{130}, 275--281 (1997).

\bibitem{Probstein_1989}
R.F. Probstein, \textit{Physicochemical Hydrodynamics}, Butterworths (1989).

\bibitem{Johnston_2023}
J. Johnston \textit{et al.}, ``A reduced-order model of concentration polarization in reverse osmosis systems with feed spacers,'' \textit{J. Membr. Sci.} \textbf{675}, 121508 (2023).

\bibitem{Geraldes_2006}
V. Geraldes and M.D. Afonso, ``Generalized mass-transfer correction factor for nanofiltration and reverse osmosis,'' \textit{AIChE J.} \textbf{52}, 3353--3362 (2006).

\bibitem{Wang_2021} 
L. Wang, T. Cao, J.E. Dykstra, S. Porada, P.M. Biesheuvel, and M. Elimelech, ``Salt and water transport in reverse osmosis membranes: beyond the solution-diffusion model,'' \textit{Environ. Sci. Technol.} \textbf{55}, 16665--16675 
(2021).

\bibitem{Biesheuvel_2023}
P.M. Biesheuvel, S.B. Rutten, I.I. Ryzhkov, S. Porada, and M. Elimelech, ``Theory for salt transport in charged reverse osmosis membranes: Novel analytical equations for desalination performance and experimental validation,'' \textit{Desalination} \textbf{557}, 116580 (2023).

\bibitem{Biesheuvel_2022} 
P.M. Biesheuvel, S. Porada, M. Elimelech, and J.E. Dykstra, ``Tutorial review of reverse osmosis and electrodialysis,'' \textit{J. Membr. Sci.} \textbf{647}, 120221 (2022). % arXiv:2110.07506 (2021).

\bibitem{Junker_2021}
M.A. Junker, W.M. de Vos, R.G.H. Lammertink, and J. de Grooth, ``Bridging the gap between lab-scale and commercial dimensions of hollow fiber nanofiltration membranes,'' \textit{J. Membr. Sci.} \textbf{624}, 119100 (2021).

\bibitem{Hamer_Wu_1972}
W.J. Hamer and Y.-C. Wu, ``Osmotic coefficients and mean activity coefficients of uni-univalent electrolytes in water at 25 $^\circ$C,'' \textit{J. Phys. Chem. Ref. Data} \textbf{1}, 1047--1099 (1972). % https://doi.org/10.1063/1.3253108  1047-1099

\bibitem{Blankert_2022} 
B. Blankert, K.Th. Huisman, F.D. Martinez, J.S. Vrouwenvelder, and C. Picioreanu, ``Are commercial polyamide seawater and brackish water membranes effectively charged?'' \textit{J.~Membr. Sci. Lett.} \textbf{2}, 100032 (2022).

\bibitem{Bouchard_1994}
Ch.R. Bouchard, P.J. Carreau, T. Matsuura, and S. Sourirajan, ``Modeling of ultrafiltration: predictions of concentration polarization effects,'' \textit{J. Membr. Sci.} \textbf{98}, 215--229 (1994).

%\bibitem{Denisov_1994} G.A. Denisov, ``Theory of concentration polarization in cross-flow ultrafiltration: gel-layer model and osmotic-pressure model,'' \textit{J. Membrane Sci.} \textbf{91}, 173--187 (1994).

%\bibitem{Starov_1993} V.M. Starov and N.V. Churaev, ``Separation of electrolyte solutions by reverse osmosis,'' \textit{Adv.~Colloid Interface Sci.} \textbf{43}, 145--167 (1993).

%\bibitem{Pranic_2021} M. Prani\'{c}, E.M. Kimani, P.M. Biesheuvel, and S. Porada, ``Desalination of Complex Multi-Ionic Solutions by Reverse Osmosis at Different pH Values, Temperatures, and Compositions,'' \textit{ACS Omega} \textbf{6}, 19946--19955 (2021).

%\bibitem{Jonnson_1980} G. Jonsson, ``Overview of theories for water and solute transport in UF/RO membranes,'' \textit{Desalination} \textbf{35} 21--38 (1980).  %https://doi.org/10.1016/S0011-9164(00)88602-6

%\bibitem{Bowen_2002}  W.R. Bowen and J.S. Welfoot, ``Modelling the performance of membrane nanofiltration -- critical assessment and model development,'' \textit{Chem. Eng. Sci.} \textbf{57}, 1121--1137 (2002).
	
%\bibitem{Verbrugge_1990} M.W. Verbrugge and R.F. Hill, ``Ion and solvent transport in ion-exchange membranes. I. A macrohomogeneous mathematical model,'' \textit{J. Electrochem. Soc.} \textbf{137}, 886--893 (1990).		
	
%\bibitem{Oren_2018} Y.S. Oren and P.M. Biesheuvel, ``Theory of ion and water transport in reverse osmosis membranes,'' \textit{Phys Rev. Appl.} \textbf{9}, 024034 (2018).
	
%\bibitem{Biesheuvel_2020} P.M. Biesheuvel, L. Zhang, P. Gasquet, B. Blankert, M. Elimelech, and W.G.J. van der Meer, ``Ion selectivity in brackish water desalination by reverse osmosis: theory, measurements, and implications,'' \textit{Environ. Sci. Technol. Lett.} \textbf{7}, 42--47 (2020).

%\bibitem{Wijmans_1985} J.G. Wijmans, S. Nakao, J.W.A. van den Berg, F.R. Troelstra, and C.A. Smolders, ``Hydrodynamic resistance of concentration polarization boundary layers in ultrafiltration,'' \textit{J. Membr. Sci.} \textbf{22}, 117--135 (1985).

%\bibitem{Armstrong_2022} Armstrong Riley Coronell Trends and errors in reverse osmosis membrane performance  calculations stemming from test pressure and simplifying assumptions about concentration polarization and solute rejection (2022)

%\bibitem{Malek_1996} A. Malek, M.N.A. Hawlader, and J.C. Ho, ``Design and economics of RO seawater desalination,'' \textit{Desalination} \textbf{105}, 246--261 (1996).
	
%\bibitem{Ong_2002} S.L. Ong, W. Zhou, L. Song, and W.J. Ng, ``Evaluation of feed concentration effects on salt/ion transport through RO/NF membranes with the Nernst-Planck-Donnan model,'' \textit{Env. Eng Sci.} \textbf{19}, 429--439 (2002)

% \bibitem{Tiraferri_2022} A. Tiraferri, M. Malaguti, M. Mohamed, M. Giagnorio, F.J. Aschmoneit, ``A new algebraic water flux equation for the streamlined evaluation and design of membranes and membrane systems,'' (2022).

%\bibitem{Biesheuvel_Dykstra_2020} P.M. Biesheuvel and J.E. Dykstra, \textit{Physics of Electrochemical Processes},   ISBN:9789090332581 (2020). % http://www.physicsofelectrochemicalprocesses.com 

% \bibitem{Chmiel_2006} H. Chmiel, X. Lefebvre, V. Mavrov, M. Noronha, and J. Palmeri, ``Computer Simulation of Nanofiltration, Membranes and Processes,'' in: Handbook of Theoretical and Computational Nanotechnology, Ed. M. Rieth, W. Schommers, \textbf{5}, 93--214, Amer. Sci. Publ. (2006). %American Scientific Publishers (2006). % er is ook sprake van 2005
	
\end{thebibliography}
\end{document}